\documentclass[twoside]{article} 

\usepackage[accepted]{aistats2017}
\usepackage{framed}
\usepackage{graphicx} 
\usepackage{subfigure} 
\usepackage{natbib}
\usepackage{algorithm}
\usepackage{amssymb,amsmath,dsfont}
\usepackage[svgnames]{xcolor}

\newcommand{\bs}[1]{\boldsymbol{#1}}
\newcommand{\mc}[1]{\mathcal{#1}}
\newcommand{\mr}[1]{\mathrm{#1}}
\newcommand{\bm}[1]{\mathbf{#1}}
\newcommand{\ds}[1]{\mathds{#1}}

\begin{document}


\runningtitle{Scalable semiparametrics for heavy tails}

%

\twocolumn[

\aistatstitle{Scalable semiparametric inference for the \\means of heavy-tailed distributions}

\aistatsauthor{
  Matt Taddy \And Hedibert Freitas Lopes \And Matthew Gardner}

\aistatsaddress{ 
  Microsoft Research and Chicago Booth \And 
  Insper S\~ao Paulo \And eBay Inc.
}
]


\begin{abstract} 
Heavy tailed distributions present a tough setting for inference. They are also common in industrial applications, particularly with Internet transaction datasets, and machine learners often analyze such data without considering the biases and risks associated with the misuse of standard tools. This paper outlines a procedure for inference about the mean of a (possibly conditional) heavy tailed distribution that combines nonparametric analysis for the bulk of the support with Bayesian parametric modeling -- motivated from extreme value theory -- for the heavy tail. The procedure is fast and massively scalable.  The resulting point estimators attain lowest-possible error rates and, unique among alternatives, we are able to provide accurate uncertainty quantification for these estimators. The work should find application in settings wherever correct inference is important and reward tails are heavy; we illustrate the framework in causal inference for A/B experiments involving  hundreds of millions of users of eBay.com.
\end{abstract}

\section{Introduction}

\begin{figure*}
\centering
\begin{minipage}{.47\textwidth}
  \centering
  \includegraphics[width=\textwidth]{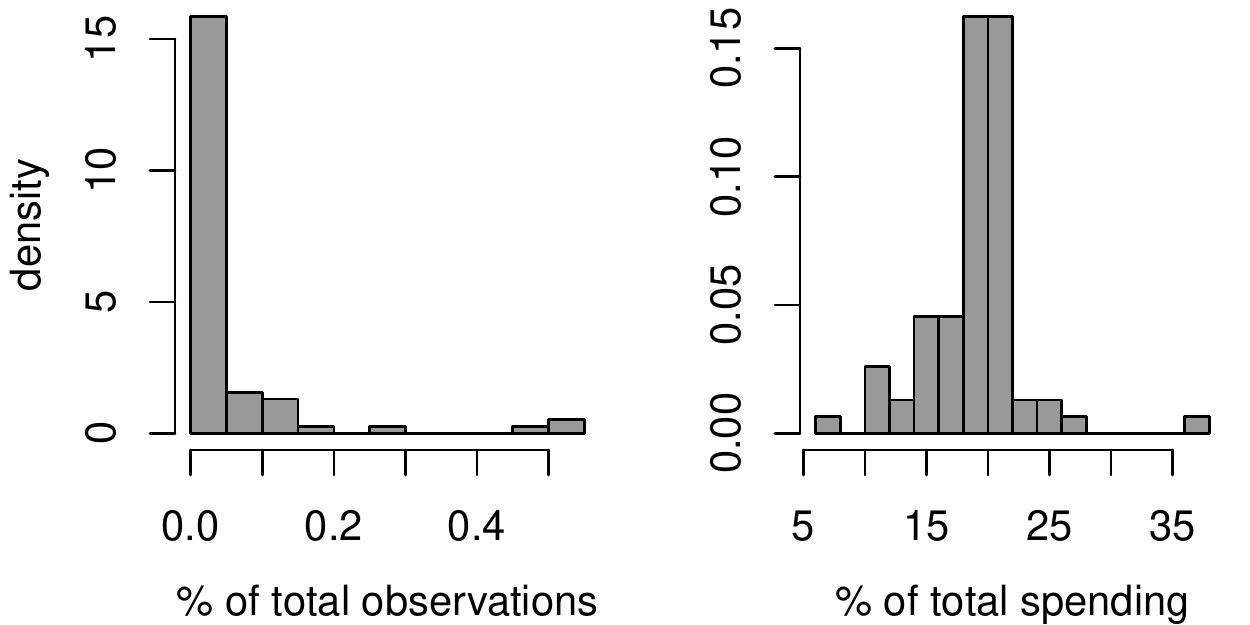}
  \caption{\label{fig:proportions}  The proportion of observations (left) and of total spending (right) due to users spending greater than \$2000 in each treatment group. }
\end{minipage}\hfill
\begin{minipage}{.47\textwidth}
  \centering
  \includegraphics[width=\textwidth]{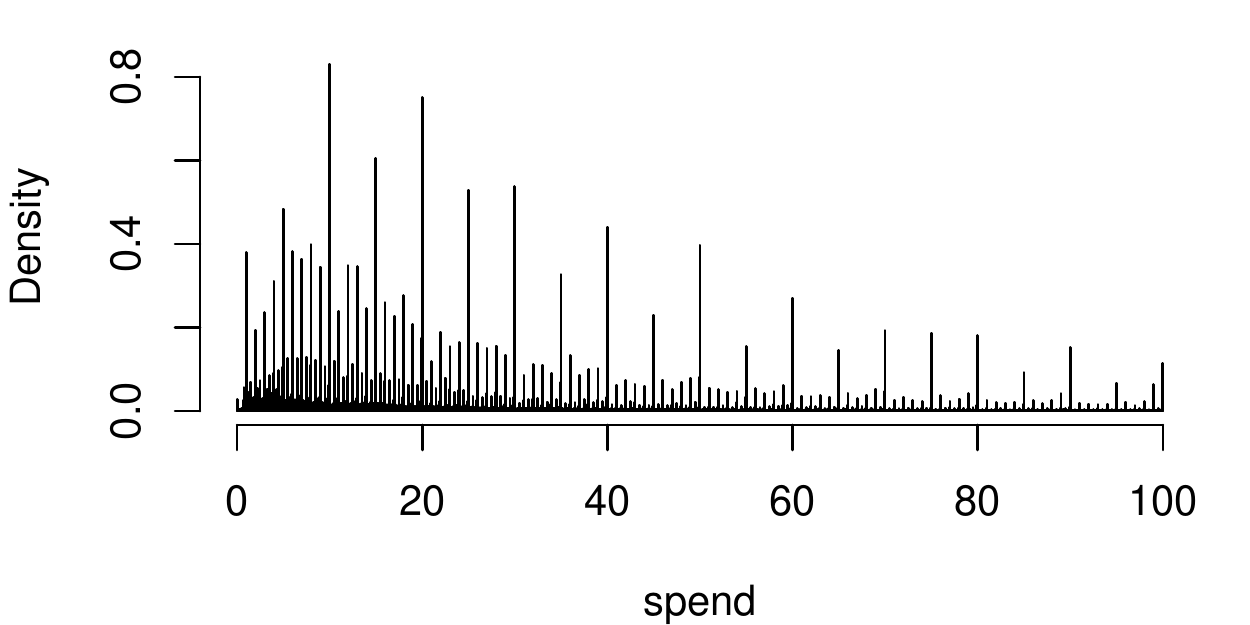}
  \caption{\label{fig:lowval} Sample distribution for user spending values below \$100 in the experiments of Section \ref{sec:perf}.}
\end{minipage}
\end{figure*}

A data generating process (DGP) is {\it heavy tailed} when the
 distribution on exceedances above extreme thresholds  cannot be bounded by an
 exponential distribution.  Heavy tails are  common in measures of user 
activity on the internet
 \citep{fithian2015semiparametric,taddy_heterogeneous_2015}.  For example,
 Figure \ref{fig:proportions} illustrates spending, in US\$ per week of bought
 merchandise, across 174 sets of users on eBay.com.  Each
 sample,\footnote{These 
are targeted user subsets from past traffic.  They are not representative of
 eBay's aggregated revenue.} of 1 to 30 million users,
 corresponds to a treatment group in an A/B experiment.  In our modal
 treatment group,  less than 0.05\% of users spend more than
 \$2000; however, these users account for 20\% of the total spending.

Heavy tails imply that observations in high percentiles are  high  variance
and will have a large influence on sample means. In some cases, including 2/3
of the groups we study here, the tail variance is infinite: the fitted
parametric tail models imply nonexistent second moments.  Even when the
variance is merely near-infinite, these heavy tails have important
consequences for our inference. 
\begin{itemize} \item The learning rate for
mean inference is slower than $\sqrt{N}$, for sample size $N$, and the usual
standard error estimators tend to underestimate uncertainty
\citep[e.g.,][]{romano1999subsampling}. 
\item Gaussian error assumptions are
invalid both in finite samples and asymptotically \citep{feller1971}. 
\item
Nonparametric bootstrap estimators of sampling uncertainty about the mean will
fail: they are inconsistent for the true sampling distribution
\citep{athreya1987bootstrap}. 
\end{itemize} 
These issues have real practical
implications, and the over-sized influence of large observations on the sample
mean is well recognized by practitioners who measure on-line transactions
(e.g., when evaluating the treatment effect from an A/B trial).  A common ad-
hoc solution is to use Winsorization \citep{dixon1960}  wherein values above a
threshold are replaced by that threshold.  However, estimation is then very
sensitive to the Winsorization threshold and, due to the inconsistency of the
nonparametric bootstrap, there are no tools available for its optimal
selection or for uncertainty quantification.   At the same time, fully parametric modeling is impractical because the transaction distributions defy  summarization.  Figure \ref{fig:lowval} shows that, at the low end of the spending range, the distribution is characterized by probability spikes at discrete price points (e.g., \$1, \$99.99) and could not be represented by any standard low-dimensional parametric family.

We resolve these issues by combining nonparametric
inference for the bulk of a distribution with parametric inference for the {\it tail above a fixed threshold}.   We give a theoretically motivated rule for choosing the threshold and demonstrate that inference is robust to choices around this rule. The result is a simple framework for scalable inference with heavy tailed data. 

We highlight number of  contributions.

{\bf Scalability:~} Our algorithms provide scalable  inference in a setting where this does not exist.  Related Bayesian approaches have been proposed before (see below), but these do not scale to even moderately sized datasets and are completely infeasible on the Internet  datasets that motivate our work.  In contrast, we require no more computation on the bulk of the data than estimation of sample means and variance, and our tail inference is available via either analytical or efficient computational approximation.  

{\bf Inference:~}   Our posterior mean estimators achieve the lowest error rates available.   Moreover, our posterior standard deviations on distribution means are an accurate measure of the {\it frequentist} standard error.  Indeed, they outperform any other available standard error estimators. There exist other good and scalable point estimators for the means of heavy tailed distributions,  but none come with reliable  uncertainty quantification (which is essential in the motivating A/B trial applications). 

{\bf Consistency:~}  It is well known \citep{athreya1987bootstrap} that the usual nonparametric bootstrap is inconsistent as an estimator for the sampling distribution of the mean of a heavy tailed distribution. We introduce a novel semiparametric bootstrap and show that it is consistent for a tail threshold that grows with the sample size.   Our Bayesian inference algorithm is closely related to this semiparametric bootstrap.

{\bf Bootstrap-based posterior sampling:~}  For inference about the tail parameters, we present a novel independence Metropolis Hastings (iMH) algorithm that samples from the posterior through adjustment of the results from a parametric bootstrap.  The algorithm is trivial to code, fast and parallelizable, and its acceptance rate is a measure of the distance between Bayesian and frequentist inference. 

{\bf Extreme value analysis:~}   We contribute two general points on Bayesian analysis of heavy tails. First, our consistency analysis provides a {\it rule for choosing the tail threshold} and in both theory and practice we find that {\it results are robust around this rule}.  The threshold can thus be conditioned upon in the posterior, leading to much simpler inference than is possible if it is treated as a random variable.  Second, we find significant gains from use of informative priors on the tail {\it index} and propose a scheme for specification based upon a larger background dataset.  Informative priors on the tail {\it scale} make little difference in comparison.

Section \ref{sec:bnp} defines our general framework, Section \ref{sec:gpd}
details the parametric tail analysis, and Section \ref{sec:theory} studies
consistency.  We  illustrate our methods in analysis of A/B experiments
involving more than 100 million users of eBay.com: Section \ref{sec:perf}
validates performance through subsampling of treatment groups, and Section
\ref{sec:abtrials} studies inference on treatment effects in A/B
trials.


\subsection{Related Literature}

A related  Bayesian approach is proposed in
\citet{nascimento2012semiparametric}: they combine a Dirichlet process mixture
 below a threshold with a generalized Pareto above.  All parameters, including the value
of the threshold itself, are sampled from their joint posterior in an MCMC
algorithm.  Unfortunately, this procedure is completely non-scalable; it takes
around 1 second {\it per posterior draw} when analyzing one of the small
subsamples from Section \ref{sec:perf}.  The continuous mixture model is also a poor fit for Internet transaction datasets, which include density spikes at discrete values (e.g., \$0.99, \$99).
 In any case, the MCMC failed to converge without tight priors on the tail threshold or scale,  and 
yields poorly performing estimators with errors larger than those from the naive sample mean. 

\citet{johansson2003estimating} describes estimation for the mean of a
heavy tailed distribution that combines the sample mean below a threshold with
the mean of a maximum likelihood  generalized Pareto above that threshold. The
point estimates from this approach are equivalent to those from our procedure under the non-informative prior with Laplace posterior approximation.   
 Johansson's asymptotic variance formulas depend upon unknown model
parameters and thus cannot be applied in practical inference.

\citet{romano1999subsampling} use without-replacement sub-sampling  to estimate the sampling distribution for the  mean of a heavy tailed sample.   We
discuss and compare to their estimators in our applications.

Finally,  \citet{fithian2015semiparametric} estimate tail distributions through exponential tilting of models fit on larger samples.
This shares with our informative-prior models a strategy of using background
datasets to inform individual tails.  Their tilting estimator works nicely: it provides point estimation that is as good as our best methods. However, there is no uncertainty quantification available in
\cite{fithian2015semiparametric}.

\section{A Semiparametric model for heavy tailed data generating processes}
\label{sec:bnp}

Our inference strategy is built around the use of  Dirichlet-multinomial
sampling  as a flexible representation for an arbitrary data generating
process (DGP). In its standard application, this model treats the observed sample
as a draw from a multinomial distribution over a large but finite set of
support points.   A Dirichlet prior is placed on the probabilities in this
multinomial, and the posterior distribution over possible DGPs is induced by
the posterior on these probabilities.   The approach has a long history.  It
was introduced by Ferguson \citep{ferguson_bayesian_1973}, it serves as the
foundation for the Bayesian bootstrap \citep{rubin_bayesian_1981}, and it has
been studied by numerous authors 
\citep{chamberlain_nonparametric_2003,lancaster_note_2003,
poirier_bayesian_2011,taddy_forests_2015,taddy_heterogeneous_2015}.

Our work presents an extension of the standard Dirichlet-multinomial
scheme. Consider a univariate random variable, say $z$. We assume the usual
fully-nonparametric model below a certain fixed \textit{threshold}, say $u$.
That is, the prior DGP for $z < u$ is a Multinomial draw, with Dirichlet
distributed probability, from  a large-but-finite number of support
points.  At the same time, with some 
probability  our realized $z$ is instead drawn as $u + v$ where $v>0$ is a
random {\it exceedance} from some distribution.  

We model our tail exceedances as realizations from a generalized Pareto
 distribution (GPD), with  $\mr{p}(V <
 v) = 1 - (1 +
\xi v /\sigma)^{-1/\xi}$ and density function on $v >0$
\begin{equation}\label{eq:gpddensity}
\mathrm{GPD}(e;~\xi,~\sigma) = 
\frac{1}{\sigma}\left( 1 + \xi \frac{v}{\sigma}\right)^{-(\frac{1}{\xi} + 1)}
\end{equation}
for tail index $\xi>0$ and scale $\sigma>0$.  The generalized Pareto is a
commonly applied tail model 
\citep{smith1989extreme,davison1990models,pickands1994bayes,
johansson2003estimating,fithian2015semiparametric} 
with  justification as the limiting distribution for exceedance
beyond large $u$ for a wide family of processes
\citep{pickands1975statistical,smith1987estimating,coles1996bayesian}.  
As $\xi \rightarrow 0$ the GPD converges to an exponential distribution, and
for $\xi > 0$ the tails are heavier-than-exponential.  For $\xi \geq 1/2$ the
variance of $v$ is infinite, and for $\xi \geq 1$ the mean is infinite.  Our analysis focuses on 
$\xi \in (0,1)$, so that the tail
is heavy enough to cause problems  but not so heavy that the mean does not exist.

Combining the GPD and Dirichlet-multinomial 
sampling yields our semi-parametric model,
\begin{equation}\label{eq:fullmod}
g(z) = \frac{1}{|\bs{\theta}|}
\sum_{l=1}^{L} \theta_l \ds{1}_{[z =
\zeta_l]} + \frac{\theta_{L+1}}{|\bs{\theta}|}  \mathrm{GPD}(z-u;~\xi,~\sigma)\ds{1}_{[z \geq u]}
\end{equation} where $\bs{\mc{Z}} = \{\zeta_1 \dots
\zeta_{L}\}$, all elements less than $u$, is the support for the bulk of the DGP $g(z)$,  $\bs{\theta} = [\theta_1 \cdots \theta_{L+1}]'$ is a vector of
random weights with $\theta_l\geq 0 \;\forall  \;l$, and $|\bs{\theta}| = \sum_i |\theta_i|$.

Observations are assumed drawn
independently from (\ref{eq:fullmod}) by first sampling $l_i$
with probability $\theta_{l_i}$ and then assigning $z_i =
\zeta_{l_i}$ for $l_i \leq L$ and otherwise drawing $z_i-u \sim \mathrm{GPD}$.   A posterior over $g$ is induced by the posterior over the model parameters:
$\bs{\theta}$, $\xi$, and $\sigma$.  Functionals of $g$, such as $\ds{E}_gf(z)$ for arbitrary
function $f$ and where $\ds{E}_g$ implies expectation over $z\sim g$, are
thus random variables.

\subsection{Inference on the sampling weights}

A conjugate prior places independent
exponential distributions on each weight: $\theta_l\sim
\mr{Exp}(a)$ for $l=1\dots L+1$, where $\ds{E}[\theta_l] = a$ and $a>0$ is the prior `rate'.\footnote{This is equivalent to a Dirichlet distribution on  normalized
weights, $\bs{\theta}/|\bs{\theta}|$. 
} After observing a sample $\bm{Z} = [z_1
\cdots z_{N}]'$, each weight remains independent in the posterior with
 distribution $\theta_l|\bm{Z} \sim \mr{Exp}(a +
\sum_{i=1}^{N}\ds{1}_{[l_i=l]})$.
We  focus on
the limiting prior that arises as $a\rightarrow 0$ \citep{rubin_bayesian_1981,chamberlain_nonparametric_2003,taddy_forests_2015,taddy_heterogeneous_2015}.  This `non-informative'
limit yields a massive computational convenience: as $a\rightarrow 0$ the
weights for unobserved support points converge to a point mass
at zero: $p(\theta_l = 0 |\bm{Z}) = 1$ if $l \neq l_i$ $\forall i$. Our posterior is then a multinomial sampling model with random positive weights on only the \emph{observed data points} and on the tail ($l_i = L+1$). 

To simplify notation, say $z_i < u$ for $i\leq m$ and $z_i \geq u$ for $i=m+1,\dots,m+n$ with $N=m+n$.   We then overload  and
re-write $\bs{\theta} = [\theta_1,
\dots, \theta_m, \theta_{m+1}]'$ as the posterior vector of 
weights on observations $z_1, \dots, z_m$ (all less than $u$; repeated values are fine) and on the tail.
A posterior DGP realization is 
\begin{align}\label{dgppost}
 g(z) \mid \bm{z}, \xi, \sigma = &\frac{1}{|\bs{\theta}|}\sum_{i=1}^m \theta_i \ds{1}_{[z =
z_i]} + \notag \\ &+\frac{\theta_{m+1}}{|\bs{\theta}|}  \mathrm{GPD}(z-u;~\xi,~\sigma)\ds{1}_{[z \geq u]},
 \end{align}
 with $\theta_i \stackrel{iid}{\sim} \mr{Exp}(1) ~\forall i \leq m~~\text{and}~~ \theta_{m+1} \sim \mr{Exp}(n)$.
Details on the GPD tail posterior  are in Section \ref{sec:gpd}.

\subsection{Inference on the DGP Mean}

The mean of $g(z)$ is the random variable 
$\mu = \ds{E}z
=\left(\sum_{i=1}^m \theta_i z_i 
+ \theta_{m+1} ( u + \sigma(1-\xi)^{-1}\right)/{|\bs{\theta}|}.$
Uncertainty about $\ds{E}z$ is induced by the posterior on 
$\bs{\theta}$ and on the mean exceedance $\lambda = \sigma/(1-\xi)$. Because
$u$ is fixed, we have $\bs{\theta} \perp\!\!\!\perp\lambda$. 
It is easy to
see that $\ds{E}\mu = \tfrac{1}{m+n}\sum_{i=1}^m z_i + \frac{n}{m+n}(u +
\ds{E}\lambda)$.

The law of total variation yields posterior variance
$
\mr{var} \mu = \ds{E}[\mr{var}(\mu|\lambda)] + \mr{var}(\ds{E}[\mu|\lambda]).
$ Given  properties of the Dirichlet posterior on $\bs{\theta}/|\bs{\theta}|$, the first term is
\begin{align}
\ds{E}[\mr{var}(\mu|\lambda)] 
 = &\frac{\sum_{i=1}^{m} (z_i - \ds{E}\mu)^2 + n(u+\ds{E}\lambda - \ds{E}\mu)^2}{(m+n)(m+n+1)} \notag\\&+ \frac{n^2 (m+n-2)\mr{var}(\lambda)}{(m+n)^2(m+n+1)}
\end{align}
where $\mu_{\lambda} = \left[\sum_{i=1}^m z_i + n(u+\lambda)\right]/(m+n)$ and $\ds{E}\mu$ is the posterior expectation from above.  The second term is
$
\mr{var}(\ds{E}[\mu|\lambda]) = \frac{n^2}{(m+n)^2} \mr{var}(\lambda)
$
 and the full expression
\begin{align}
\mr{var} \mu = &
\frac{\sum_{i=1}^{m} (z_i - \ds{E}\mu)^2 + n(u+\ds{E}\lambda - \ds{E}\mu)^2}{(m+n)(m+n+1)} \notag\\&+ \frac{2n^2 (m+n-0.5)}{(m+n)^2(m+n+1)}\mr{var}(\lambda).
\end{align}
Noting that $\lambda = \sigma/(1-\xi)$, the necessary tail moments $\ds{E}\lambda$ and $\mr{var}(\lambda)$ are available through either
Laplace approximation or MCMC as described below.

\section{Inference for tail parameters}
\label{sec:gpd}

 In this section we describe Bayesian modeling and inference for the GPD parameters, $\xi$ and $\sigma$, conditional upon the sample of exceedances $\{ v_i = z_{m+i}-u \}_{1}^n$. We are focusing on heavy tails with finite mean exceedances that correspond to $\xi \in (0,1)$.  On this range, $\sigma$ can take any positive value.  A simple independent prior setup is then
$\pi(\sigma,\xi) = \mr{Beta}(\xi; a,b)\mr{Ga}(\sigma; c, d) \propto \xi^{a-1}(1-\xi)^{b-1} \sigma^{c-1}e^{-d\sigma}$, where $\mr{Beta}(\cdot~;~a,b)$ denotes a beta density with mean $a/(a+b)$ and $\mr{Ga}(\cdot~;~c,d)$ a gamma density with mean $c/d$, with $a,b,c,d >0$.  We work primarily with a version of this prior  takes the limit $c,d \rightarrow 0$ to obtain
\begin{equation}\label{eq:gpdrefprior}
\pi(\sigma,\xi) = \frac{1}{\sigma}\xi^{a-1}(1-\xi)^{b-1}\mathds{1}_{\xi \in (0,1)},
\end{equation}
the combination of a beta on $\xi$ and an improper uniform prior on $\log\sigma$.
Following results in \cite{northrop2015posterior} and
\cite{castellanos2007default}, the posterior for $[\sigma,\xi]$ will be
proper under this prior given a minimum of three
observations.

Our beta-gamma prior combines with the GPD likelihood to yield a log posterior proportional to
$
l(\sigma,\xi)=-\frac{1+\xi}{\xi}\sum_i
\log\left(1 + \xi \frac{v_i}{\sigma}\right)
+ (a\!-\!1)\log\xi + (b\!-\!1)\log(1\!-\!\xi) + (c\!-\!n\!-\!1)\log\sigma - d\sigma.
$
Maximization of this objective leads to MAP estimates of the parameters, say $[\hat \xi, \hat \sigma]$.  The related problem of MLE estimation for GPDs is well studied by \citet{grimshaw1993computing} and his algorithm is easily adapted for fast MAP estimation within our domain $[\xi,\sigma] \in (0,1)\times \ds{R}^+$.

\subsection{Laplace posterior approximation}

For fast approximate inference, this section proposes analytic posterior approximation via Laplace's method  centered on the posterior mode.
The main object of interest is the posterior for the GPD mean, $\sigma/(1-\xi)$.
We make the transformation
\begin{align}
\lambda = \frac{\sigma}{1-\xi} \Leftrightarrow \sigma = \lambda(1-\xi),
\end{align}
with inverse Jacobian $|J| = 1-\xi$, to obtain the posterior
\begin{align}
\mr{p}(&\lambda,\xi \mid \mathbf{v} ) \propto \\& \frac{\xi^{a-1}e^{-d\lambda(1-\xi)}}{\lambda^{n-c+1}(1-\xi)^{n-b-c+1}}
\prod_i \left(1 + \frac{\xi}{1-\xi}\frac{v_i}{\lambda}\right)^{-\left(\frac{1}{\xi}+1\right)}.\notag
\end{align}
Note that the MAP estimate for $\lambda$ is just $\hat\lambda =
\hat\sigma/(1-\hat\xi)$. The Laplace approximation
\citep{tierney_accurate_1986} to the \textit{marginal} posterior distribution
on $\lambda$  is available as
$
\mathrm{\hat p}\left( \lambda \mid \bm{x} \right) 
= \mathrm{N}\left( \hat\lambda, -\left.\nabla_{\lambda\lambda}^{-1}\right\vert_{\left[\hat\lambda,\hat\xi\right]}\right),
$
where $\nabla_{\lambda\lambda}$ is the curvature of the log posterior
  with respect to $\lambda$ via 
  $\displaystyle 
 \frac{\partial \log \mr{p}(\lambda,\xi |\mathbf{v})}{\partial \lambda}
= $ $$\nabla_{\lambda} = \frac{1}{\lambda}\left[(1/\xi+1)\sum_i q_i-n + c - 1\right] - d(1-\xi) $$ where $q_i = \xi v_i /\left[(1-\xi)\lambda + \xi v_i\right]$).  The approximate  variance for $\lambda$ is, with $\hat q_i = \hat\xi v_i /\left[(1-\hat\xi)\hat\lambda + \hat\xi v_i\right]$,
$$\widehat{\mr{var}}(\lambda\mid \bm{x}) 
= -\!\hat\lambda^2\!
\left[ n-\!c+\!1 \!+\! \left(\frac{1}{\hat\xi}\!+\!1\right)\sum_i \left(\hat q_i^2 \!-\! 2\hat q_i\right)\right]^{\!-1}\!\!\!.$$

\subsection{Posterior sampling and approximation}

For full posterior inference, we propose a novel independence Metropolis Hastings (iMH) algorithm \citep[e.g.,][]{gamerman_markov_2006}
that uses a parametric bootstrap of the MAP estimates as an MCMC proposal distribution.  This approach  is similar to the bootstrap reweighting of \citet{efron2012bayesian}, but unlike that work it does not require an analytic expression for the sampling distribution of the statistics of interest.

\begin{algorithm}
{\bf Bootstrap iMH posterior sampler}
\begin{itemize}
\item Fit the MAP parameter estimates $[\hat\xi,\hat\sigma]$ to maximize the log posterior objective $l(\xi, \sigma)$.
\item Draw  $\{\hat\xi^b,\hat\sigma^b \}_{b=1}^B$ from the parametric bootstrap:
\begin{itemize}
\item Generate a sample $\{z_i^b\}_{i=1}^n$  by simulating from the MAP estimated model $\mr{GPD}(\hat\xi,\hat\sigma)$.
\item Obtain new MAP estimates $[\hat\xi^b,\hat\sigma^b]$ conditional upon $\{z_i^b\}_{i=1}^n$.
\end{itemize}
\item Estimate the  bootstrap distribution, say $r(\xi,\sigma)$, via kernel smoothing on $\{\hat\xi^b,\hat\sigma^b \}_{b=1}^B$.
\item For $b=2 \dots B$, replace $[\hat\xi^b,\hat\sigma^b]$ with $[\hat\xi^{b-1},\hat\sigma^{b-1}]$ with probability 
\[
1 - \mr{min}\left\{ 
\frac{r(\hat\xi^{b-1},\hat\sigma^{b-1})\exp[l(\hat\xi^b, \hat\sigma^b)]}
{r(\hat\xi_{b},\hat\sigma_{b})\exp[l(\hat\xi^{b-1}, \hat\sigma^{b-1})]}, ~1\right\}.
\]
\end{itemize}
\end{algorithm}

This  is simple and fast.   It also connects Bayesian and frequentist inference: high acceptance rates imply a posterior close to the sampling distribution.   We emphasize that this is a novel recipe for generating MCMC algorithms from bootstrap samples, and due to the often close relationship between sampling distributions and posteriors we expect that this recipe will be useful in a wide variety of additional settings.

\subsection{Background tails and informative priors}
\label{sec:prior}

It is common to expect similar tail properties across multiple distributions.  For example, we believe that small changes to the eBay website have negligible effect on whether a user makes a big purchase.  This information can be used in a prior that shrinks each tail towards a larger background dataset.

We focus on adding information on the tail index, $\xi$, under the prior in (\ref{eq:gpdrefprior}).  The tails of related distributions tend to converge to a GPD with the same index \citep{pickands1975statistical} and there is abundant precedence for analysis of multiple  distributions using a shared tail index \citep{davison1990models,fithian2015semiparametric}.  
If you believe that every group has the {\it same} tail index, use as your prior for $\xi$ the posterior from analysis of a larger dataset.   Applying the methods of Section \ref{sec:gpd} to  100,000 users with spending over \$2000, we obtain a posterior, and hence prior, on $\xi$ that is well approximated by a $\mr{Beta}(80,80)$ distribution.
Alternatively, if you believe that each treatment group has a different-but-similar tail index, specify the $\mr{Beta}(a,b)$ distribution that best fits a sample of estimated tail indexes from prior analyses. In our eBay example, considering a set of 149 $\hat \xi$ from samples not analyzed in Sections \ref{sec:perf} --\ref{sec:abtrials}, this yields a $\mr{Beta}(9,9)$ prior.

In Section \ref{sec:perf} we find that both priors -- the hierarchical-model
$\mr{Beta}(9,9)$ and the single-background-tail $\mr{Beta}(80,80)$ -- lead to
significant improvements in estimation.  In
contrast,  we generally do not recommend using an informative prior on $\sigma$.   The theory of Section
\ref{sec:theory} implies that this scale parameter is sensitive to
$u$, so that information at one threshold does not inform estimation at another.   Experimentation  under
the setup of Section \ref{sec:perf} for priors on $\sigma$ (not shown) generally led to higher errors
and lower ratios of the posterior standard deviation relative to the true
error rate.

\section{Consistency of the semiparametric heavy tailed bootstrap}
\label{sec:theory}

Consider inference
about $Q_N = \sqrt{N}(\hat\mu_N - \mu)$ for a sample of $N$ observations drawn
from true distribution function $F(z)$, with $\int_0^\infty z dF(z) = \mu <
\infty$ and where $\hat \mu_N$ denotes the  MLE of $\mu$ for a
size-$N$ sample from $F$.  A  bootstrap replaces $F \approx \widehat
F_N$ and uses this to obtain $b=1,\ldots,B$ draws of $Q_N^b =
\sqrt{N}(\hat \mu_N^b - \hat \mu_N)$ where $\hat \mu_N^b$ is the MLE for a 
size-$N$ sample from $\widehat F_N$. The targeted sampling distribution,
 $G_{N}(q) = \mr{p}(Q_N < q)$, is  estimated as $\widehat G_{N}(q) =
 B^{-1}\sum_{b=1}^B  \ds{1}_{[Q^b_N <q]}$.

Standard  results \citep{bickel1981some,beran2003impact} require that $\widehat G_N$ converges in distribution  to $ G_\infty$ \textit{uniformly} across all $\widehat F_N$ in a neighborhood, say $\mc{F}$,  containing $F$ and also $\widehat F_N$ for $N$ big enough (in addition, the mapping $F \mapsto G_{\infty}$ must be continuous).  Convergence in probability for $\widehat F_N(z)$ to $F(z)$ $\forall z$ then implies consistency of $\widehat G_{N}$ in that, as $N\to\infty$, $\mr{p}( |\widehat G_N(q) - G_N(q)| < \epsilon ) \to 0$ for all $q$ and $\epsilon>0$.

\citet{athreya1987bootstrap} shows that the nonparametric bootstrap -- using the empirical distribution function (EDF) as $\hat F_N$ -- is inconsistent for the distribution of the sample mean for data that has infinite variance. As explained  by \citet{hall1990asymptotic}, in this setting $\widehat G_N$ based upon samples from $\hat F_N$ does not converge uniformly to $G_\infty$ because sums of the largest re-sampled observations, $\sum_{i=N-r}^N z^b_{(i)}$ for $r \geq 1$, can be  dominated by repeats of the largest sample observation, $z_{(N)}$.

Instead,  define a {\it semiparametric bootstrap} that is the  
frequentist analogue of our Bayesian procedure. 

\begin{algorithm}

{\bf Semiparametric Frequentist Bootstrap}

\vskip .25cm
Given MLE parameters, $[\hat \xi_n, \hat \sigma_n]$,  for $b=1,\ldots,B$:
\begin{itemize}
	\item draw $m_b \sim \mr{Bin}(m/N, N)$ and set $n_b = N-m_b$;
	\item sample with replacement $m_b$ observations from 
	$\{z_i: z_i < u\}$, say $\{z^b_{1}, \ldots, z^b_{m_b}\}$;
	\item draw $v^b_1 \ldots v^b_{n_b}$ from $\mr{GPD}(\hat \xi_n, \hat \sigma_n)$ and fit the corresponding MLE, 
$\hat\lambda^b_{n_b} = \hat\sigma^b_{n_b}/(1-\hat \xi^b_{n_b})$; 
\item   set $\hat \mu^{b}_N = \left( \sum_{i=1}^{m_b} 
z_{i_b} + n_b (u+\hat\lambda^b_{n_b})\right)/N$.   
\end{itemize}
\end{algorithm}

The sampling distribution, e.g., for $\sqrt{N}(\hat\mu_N - \mu)$ is then approximated by $\left\{\sqrt{N}(\hat\mu_N^b - \hat\mu_N)\right\}_{b=1}^B$.

This semiparametric bootstrap is the combination of three bootstrap
estimators, for distributions on  $\frac{1}{m}\sum_{i=1}^{N}
z_{i}\ds{1}_{[z_i<u]}$, on $m/N$, and on $\hat\lambda_{n}$.  Consistency of
the nonparametric bootstrap for the first two statistics is established through
standard arguments
\citep{Mammen1992}. Therefore, to show consistency for the semiparametric bootstrap we need only to confirm that the  bootstrap using $\widehat
F_N(z-u|z\geq u) = \mr{GPD}(\hat \xi_n, \hat \sigma_n)$ converges to the correct
distribution for $\hat \lambda_n$.

\citet{johansson2003estimating} considers DGPs with distribution functions
$F(z) = 1 -cz^{-1/\zeta}(1 + z^{-\delta}L(z))$, where $c,\delta > 0$ and
$L(tz)/L(z) \to 1$ with $z\to \infty$ for $t>0$.  This defines a wide class of
heavy tailed distributions, and for $u_N$ large enough the distribution
$F(z-u_N |z \geq u_N)$ approaches a $\mr{GPD}(\xi, \sigma_N)$ where $\sigma_N
= u_N \xi$.  Following the same steps as Johansson, which apply results from
\citet{smith1987estimating} on the asymptotic distribution for  MLEs
$[\hat\xi_n,\hat\sigma_n]$, you can show that for $F(z)$ with $\xi \in (0,1)$
and  $z^{-\delta}L(z)$  non-increasing, if $u_N = O(N^{\xi/(1+2\delta\xi)})$
then
\begin{equation}\label{lambdadist}
\sqrt{n}\left(\hat \lambda_n - \ds{E}_F[z-u_N|z\geq u_N]\right) \rightarrow_d \mr{N}(0, q_n)
\end{equation}
where $q_n = \hat\sigma_N(1+\xi)(1-\xi+2\xi^2)/(1-\xi)^4$.  Thus our
 bootstrap sample generator, $\mr{GPD}(\hat \xi_n, \hat \sigma_n)$, converges to 
$F(z-u_N |z \geq u_N)$ along a sequence of distributions with means $\hat
\lambda_n$ that are asymptotically normal around the target,
$\ds{E}_F[z-u_N|z\geq u_N]$.  From \citet{beran1997diagnosing}, this  establishes consistency of the tail bootstrap, and hence of our full semiparametric bootstrap.

 The bootstrap succeeds here because  MLEs
  converge quickly to the `true' GPD model; inference is then based
 upon {\it new} samples from this distribution and, unlike resamples from the
 EDF, these are not overly influenced by high order statistics in the
 original sample. From (\ref{lambdadist}),   so long as $u_N$ is growing at the right rate our true tail is converging to a GPD with $\sigma_N = \xi u_N$.  We can use this fact and the {\it estimated} ratio $\hat \sigma/(\hat \xi u)$, over a set of candidate $u$, to guide threshold selection.


\begin{figure}[htb]
\begin{center}
\includegraphics[width=.45\textwidth]{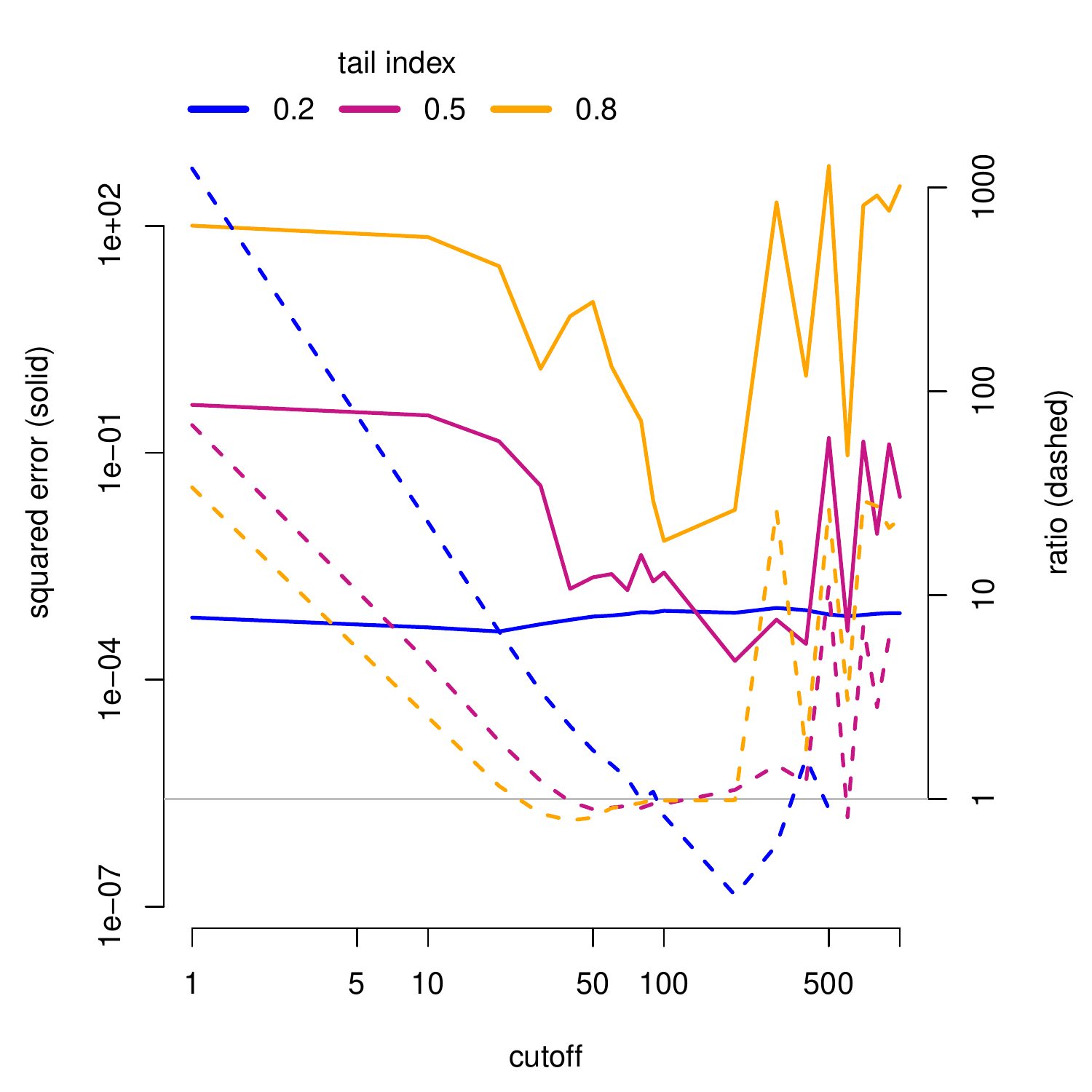}
\end{center}
\vskip -.6cm
\caption{\label{fig:fatsim} Mean estimation error (solid) and  estimated ratio  $\hat \sigma/(\hat \xi u)$ for tail thresholds $u$ and indices $\xi$. }
\end{figure}

\begin{figure*}[htb]
\begin{center}
\includegraphics[width=.9\textwidth]{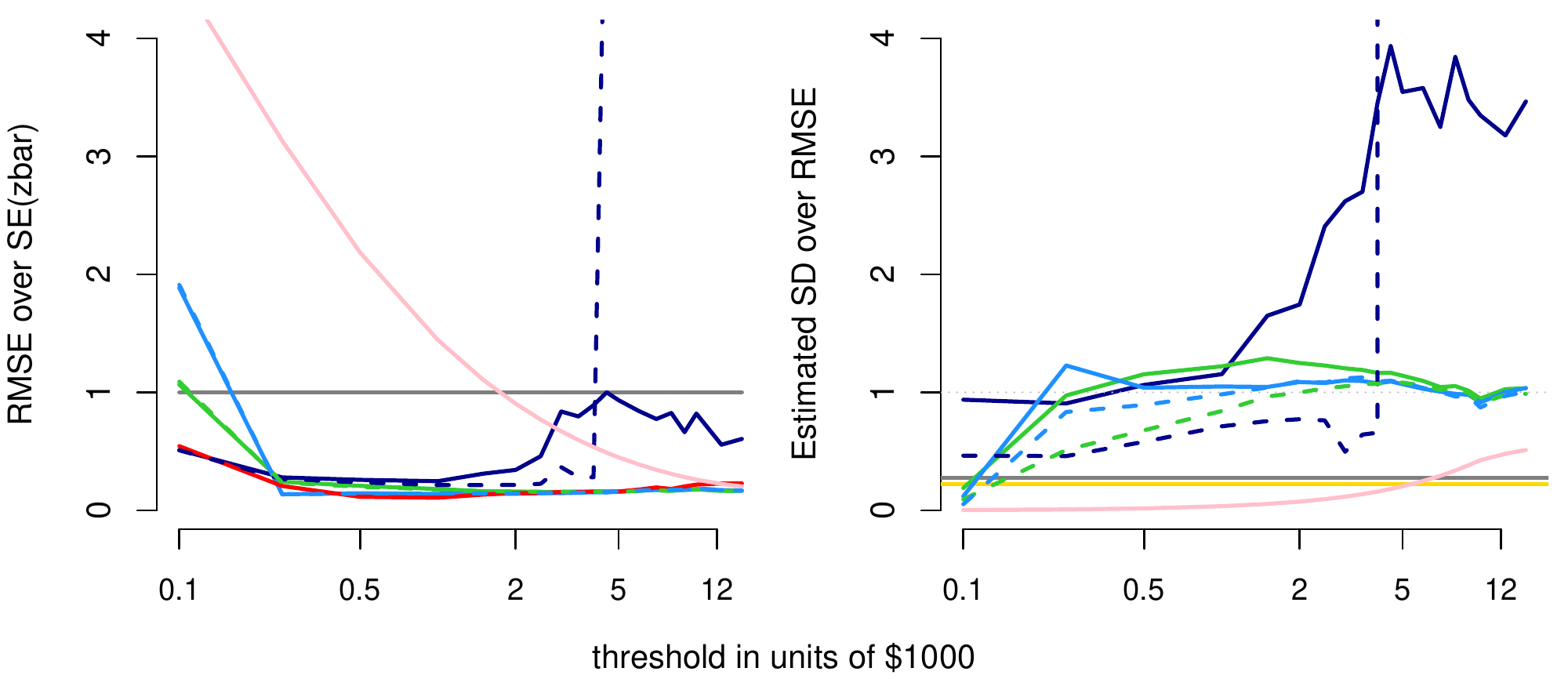}\\
{\scriptsize
\bf ~~~~~{\color{DarkBlue} -- Bayes a,b=1} \hfill {\color{DodgerBlue} -- Bayes a,b=9} \hfill{\color{LimeGreen} -- Bayes a,b=80} \hfill {\color{Pink} -- Winsorization} \hfill{\color{red} --  tilting} \hfill {\color{black!70} -- sample mean} \hfill {\color{Gold} -- N/2 subsampling} }
\end{center}
\caption{\label{fig:perf} Average  performance, over 100 samples 
of {\it N}=50,000 from each of two eBay treatment groups, as function of threshold $u$.  Our semiparametric Bayesian procedure is shown for different $\mr{Beta}(a,b)$ priors on
$\xi$, with iMH  solid
and Laplace  dashed, against results for naive sample means, Winsorized means, the tilting  of \cite{fithian2015semiparametric}, and $N/2$ subsampling standard error estimation.  The left panel shows RMSE on the full sample mean relative to performance of the naive sample mean; the right panel shows estimated standard
deviations  relative to the corresponding `true' RMSE from the left.}
\end{figure*}

\section{Simulations}
\label{sec:sim}

To illustrate our techniques and the guidance for choosing $u$, we 
study simulated data from a combination of exponential and GPD distributions.  In each simulation, we draw $10^5$ from an $\mr{Exp}(10)$ distribution and to half of these we add a draw from a $\mr{GPD}(\xi, 10)$.

We consider three tail indices, $\xi \in \{0.2,0.5,0.8\}$, that span from a near-exponential tail --  where the naive sample mean is a fine estimator -- to a heavy tail with infinite variance.  We apply our iMH semiparametric  analysis under a range of threshold values.  Results are shown in Figure \ref{fig:fatsim}.  In each case, the estimated ratio $\hat \sigma/(\hat \xi u)$ drops below one -- the value which our theory in Section \ref{sec:theory} indicates we should target -- before rising above one and becoming unstable.  In the light tailed case, the error is mostly unaffected by $u$.  For the two heavy tails -- $\xi \geq 0.5$ -- the error is lowest around the point when $\hat \sigma/(\hat \xi u)$ is equal to one and increasing in $u$.  {\it Thus, our rule  for choosing $u$ is to find the value where this ratio is near one and increasing.}

\section{Performance study}
\label{sec:perf}

Even in the presence of infinite variance and other difficulties, one can reliably measure relative performance by comparing estimators trained on \textit{small} subsamples to the corresponding full sample statistic \citep{politis1994,bickel1997resampling}.  We apply this approach on two independent eBay treatment groups, each containing more than $10^7$ observations above \$0.  For 100 repetitions on each group, we draw a subsample of $N=$50,000 and obtain, for each algorithm under study, a mean estimate based upon this subsample.  This estimate is compared to the  full-sample average, $\bar z$, and we report the discrepancy.

\begin{figure*}
\includegraphics[width=\textwidth]{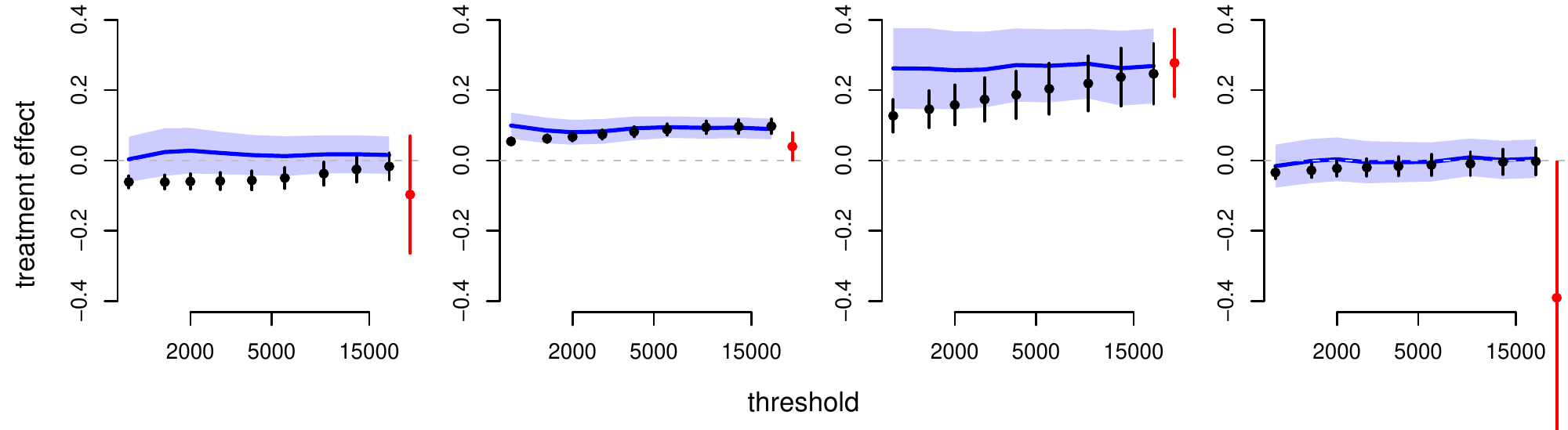}
\caption{\label{fig:semiparinf} Treatment effect estimates in four A/B trials from eBay.  
The Bayesian posterior is the region in {\bf\color{blue}blue}, Winsorized estimation is in {\bf black},   and the naive sample estimator is in {\bf\color{red}red}.  Points and lines are point estimates and intervals are $\pm1\mathrm{sd}$ or  $\pm1\mathrm{se}$, as appropriate. 
}
\end{figure*}

Resulting averages are shown in Figure \ref{fig:perf} across a range of thresholds and we make several remarks.

$\bullet$ The two semiparametric Bayesian analyses with informative priors on the tail index -- $\xi \sim \mr{Beta}(9,9)$ and $\mr{Beta}(80,80)$ -- provide superior  estimation over a wide range of thresholds $u$.  Their posterior means (from either Laplace or iMH) have the lowest or near-lowest RMSE -- around 20-40\% of the sample mean RMSE -- in both datasets for $u$ above \$500.

$\bullet$ The non-informative prior -- $\xi \sim \mr{Beta}(1,1)$ -- also leads to much lower RMSE than the sample mean for thresholds below \$4000.  However, at higher thresholds it gives larger errors than the informative prior schemes.  The iMH RMSE is still an improvement on the sample mean, but the Laplace approximation under this non-informative prior fails dramatically: RMSE explodes by an order of magnitude at high thresholds.  Since  Laplace approximation under the non-informative prior is practically equivalent to the MLE estimator of \cite{johansson2003estimating}, we see that such  techniques give terrible results for poorly chosen $u$. 

$\bullet$  The semiparametric Bayesian analyses are the {\it only} procedures that  give accurate quantification of  frequentist sampling variability. For $u>$ \$500, the two informative priors lead to posterior standard deviations that are near  the observed RMSE over our 200 subsamples.  For the $\mr{Beta}(80,80)$ prior, both Laplace and iMH standard deviations are within 10\% of the true RMSE.  The $\mr{Beta}(9,9)$ prior does worse but is still better than any alternative.  The $\mr{Beta}(1,1)$ prior with iMH sampling also provides accurate uncertainty quantification, but over the more narrow range of $u \in (\$100,\$1000)$.   In each case, our rule-of-thumb ratio $\hat \sigma/(\hat\xi u)$ is around one in these regions.

$\bullet$ For the informative priors, Laplace and iMH procedures give nearly identical RMSE for their mean estimates (the dashed and solid lines are on top of each other) but iMH standard deviations do a slightly better job replicating the observed RMSE.  As may be expected, the discrepancy between Laplace and iMH results decreases with prior information.  Also, acceptance rates on the iMH sampler were  above 90\% except at extreme thresholds, indicating that our Bayesian procedure is converging towards inference from a semiparametric frequentist bootstrap.

$\bullet$ The tilting procedure of \citet{fithian2015semiparametric}, using
the same background sample behind our $\mr{Beta}(80,80)$ prior,  yields low RMSEs at a wide range of $u$.  This  takes longer to run than 1000 iMH draws, but still finishes in seconds. Unfortunately, there is no uncertainty quantification available.

$\bullet$ Winsorization does poorly.  Its RMSE is  larger than that of the sample mean until $u>\$2000$.  The associated standard errors --  Winsorized standard deviation over $\sqrt{N}$ --  are always too small.  

$\bullet$ Unplotted, we find that use of only the GPD model (i.e., setting $u=0.01$) leads to RMSEs 5-20 times larger than that of the sample mean.   This could be predicted from the histogram in Figure \ref{fig:lowval}, which shows the sample of spend values below \$100 looking nothing like a sample from a GPD (or any continuous density).

$\bullet$  Naive standard errors for the sample mean --  $\mr{sd}(z)/\sqrt{N}$ -- are around 1/3 the observed RMSE. The subsampling standard-error estimators from \cite{romano1999subsampling}, using subsamples of size $N/2$ and estimated learning rate $N^{\min(0.5, 1-\hat\xi)}$ for MLE $\hat\xi$, lead to standard errors that are still around 70\% too small.  

Our Bayesian semiparametric procedure, especially with some  prior information for the tail index, provides lowest-possible RMSE {\it and} accurate uncertainty quantification.  Both iMH and Laplace schemes are fully scalable, but Laplace is essentially free and under the informative prior it is practically indistinguishable from the slightly more expensive iMH.

\section{A/B experiments}
\label{sec:abtrials}

Finally, we turn to the motivating application for these ideas.   In A/B experiments at eBay, two independent heavy tailed samples are obtained: one from a group receiving a treatment and another from a control group.  The object of interest is 
$
 \gamma = \mu_1 - \mu_0
$
where $\mu_1$ is the mean of the treatment group and $\mu_0$ the mean of the
control group.  The samples are independent,  so that variance on
$\gamma$ is the sum of group mean variances.

Results are shown in Figure \ref{fig:semiparinf} for four example experiments.
The Bayesian estimation here uses our informative $\mr{Beta}(80,80)$ prior and
the uncertainty bounds are based upon the Laplace
approximation. In each case,  point and uncertainty estimates for the average
treatment effects are remarkably stable across thresholds.  In contrast, 
Winsorized  estimators can change rapidly with $u$ and their standard errors are always low relative to the Bayesian
standard deviations. We also show the naive mean and standard error estimates.
In all but one case, this yields an uncertainty interval that is qualitatively
different from the Bayesian posterior; in two cases,  our semiparametric
 procedure moves the treatment effect from looking possibly
significant to insignificant, and visa versa.

\section{Conclusion}

Big Data is exciting because it allows us to estimate tiny and complicated signals. However, even with massive amounts of data you need to be careful about inference in the presence of heavy tails.  Instead of turning to a full modeling framework, which would be impractical on datasets of this size and complexity, we use simple nonparametrics for the easy bit (the middle of the distribution) while applying  careful parametric modeling on the hard bits (the tail). Although the novel iMH sampler is fast and simple (and provides a nice connection to frequentist inference), with prior information about the tail index you can avoid sampling altogether.  The procedure is massively scalable.

We have focused on single distributions (and comparisons between pairs), but the work here is applicable in many more complex modeling scenarios.  For example, any bandit learning scheme \citep[e.g.,][]{scott_modern_2010} requires accurate uncertainty quantification for the posterior distribution of rewards; when these rewards come with a heavy tail, our approach should be used.  
As another example, bagging procedures such as random forests will tend to over-fit  in the presence of extreme values
\citep{wyner2015explaining}.  Our methods can be used to
define a semiparametric loss function at leaf nodes.

\small
\bibliographystyle{plainnat}
\bibliography{../../../bib/taddy}

\begin{thebibliography}{35}
\providecommand{\natexlab}[1]{#1}
\providecommand{\url}[1]{\texttt{#1}}
\expandafter\ifx\csname urlstyle\endcsname\relax
  \providecommand{\doi}[1]{doi: #1}\else
  \providecommand{\doi}{doi: \begingroup \urlstyle{rm}\Url}\fi

\bibitem[Athreya(1987)]{athreya1987bootstrap}
KB~Athreya.
\newblock Bootstrap of the mean in the infinite variance case.
\newblock \emph{The Ann. of Stat.}, 15:\penalty0 724--731, 1987.

\bibitem[Beran(1997)]{beran1997diagnosing}
R~Beran.
\newblock Diagnosing bootstrap success.
\newblock \emph{Ann. of the Institute of Stat. Math.}, 49:\penalty0 1--24,
  1997.

\bibitem[Beran(2003)]{beran2003impact}
R~Beran.
\newblock The impact of the bootstrap on statistical algorithms and theory.
\newblock \emph{Stat. Science}, 18:\penalty0 175--184, 2003.

\bibitem[Bickel and Freedman(1981)]{bickel1981some}
P~Bickel and D~Freedman.
\newblock Some asymptotic theory for the bootstrap.
\newblock \emph{Ann. of Stat.}, 9:\penalty0 1196--1217, 1981.

\bibitem[Bickel et~al.(1997)Bickel, G{\"o}tze, and van
  Zwet]{bickel1997resampling}
PJ~Bickel, F~G{\"o}tze, and WR~van Zwet.
\newblock Resampling fewer than n observations: gains, losses, and remedies for
  losses.
\newblock \emph{Statistica Sinica}, 7, 1997.

\bibitem[Castellanos and Cabras(2007)]{castellanos2007default}
M~Eugenia Castellanos and S~Cabras.
\newblock A default {B}ayesian procedure for the generalized {P}areto
  distribution.
\newblock \emph{Journal of Statistical Planning and Inference}, 137\penalty0
  (2):\penalty0 473--483, 2007.

\bibitem[Chamberlain and Imbens(2003)]{chamberlain_nonparametric_2003}
G~Chamberlain and GW~Imbens.
\newblock Nonparametric applications of {Bayesian} inference.
\newblock \emph{Journal of Business and Economic Statistics}, 21:\penalty0
  12--18, 2003.

\bibitem[Coles and Tawn(1996)]{coles1996bayesian}
Stuart~G Coles and Jonathan~A Tawn.
\newblock A {Bayesian} analysis of extreme rainfall data.
\newblock \emph{Journal of the Royal Statistical Society. Series C (Applied
  statistics)}, pages 463--478, 1996.

\bibitem[Davison and Smith(1990)]{davison1990models}
Anthony~C Davison and Richard~L Smith.
\newblock Models for exceedances over high thresholds.
\newblock \emph{Journal of the Royal Statistical Society. Series B
  (Methodological)}, 52:\penalty0 393--442, 1990.

\bibitem[Dixon(1960)]{dixon1960}
W.~J. Dixon.
\newblock Simplified estimation from censored normal samples.
\newblock \emph{Ann. of Math. Stat.}, 31:\penalty0 385--391, 1960.

\bibitem[Efron(2012)]{efron2012bayesian}
B~Efron.
\newblock Bayesian inference and the parametric bootstrap.
\newblock \emph{Ann. of Applied Stat.}, 6\penalty0 (4):\penalty0 1971, 2012.

\bibitem[Feller(1971)]{feller1971}
W~Feller.
\newblock \emph{An Introduction to Probability Theory and Its Applications 2}.
\newblock Wiley, 2nd edition, 1971.

\bibitem[Ferguson(1973)]{ferguson_bayesian_1973}
TS~Ferguson.
\newblock A {Bayesian} analysis of some nonparametric problems.
\newblock \emph{Ann. of Stat.}, 1:\penalty0 209--230, 1973.

\bibitem[Fithian and Wager(2015)]{fithian2015semiparametric}
William Fithian and Stefan Wager.
\newblock Semiparametric exponential families for heavy-tailed data.
\newblock \emph{Biometrika}, 102:\penalty0 486--493, 2015.

\bibitem[Gamerman and Lopes(2006)]{gamerman_markov_2006}
Dani Gamerman and Hedibert~F. Lopes.
\newblock \emph{Markov {Chain} {Monte} {Carlo}}.
\newblock Chapman \& Hall/CRC, 2006.

\bibitem[Grimshaw(1993)]{grimshaw1993computing}
Scott~D Grimshaw.
\newblock Computing maximum likelihood estimates for the generalized pareto
  distribution.
\newblock \emph{Technometrics}, 35\penalty0 (2):\penalty0 185--191, 1993.

\bibitem[Hall(1990)]{hall1990asymptotic}
Peter Hall.
\newblock Asymptotic properties of the bootstrap for heavy-tailed
  distributions.
\newblock \emph{The Annals of Probability}, pages 1342--1360, 1990.

\bibitem[Johansson(2003)]{johansson2003estimating}
Joachim Johansson.
\newblock Estimating the mean of heavy-tailed distributions.
\newblock \emph{Extremes}, 6:\penalty0 91--109, 2003.

\bibitem[Lancaster(2003)]{lancaster_note_2003}
Tony Lancaster.
\newblock A note on bootstraps and robustness.
\newblock Technical report, Brown University, 2003.

\bibitem[Mammen(1992)]{Mammen1992}
Enno Mammen.
\newblock \emph{When does the bootstrap work?}
\newblock Springer, 1992.

\bibitem[Nascimento et~al.(2012)Nascimento, Gamerman, and
  Lopes]{nascimento2012semiparametric}
Fernando~Ferraz Nascimento, Dani Gamerman, and Hedibert~Freitas Lopes.
\newblock A semiparametric bayesian approach to extreme value estimation.
\newblock \emph{Statistics and Computing}, 22\penalty0 (2):\penalty0 661--675,
  2012.

\bibitem[Northrop and Attalides(2015)]{northrop2015posterior}
Paul~J Northrop and Nicolas Attalides.
\newblock Posterior propriety in {B}ayesian extreme value analyses using
  reference priors.
\newblock \emph{arXiv:1505.04983}, 2015.

\bibitem[Pickands(1975)]{pickands1975statistical}
James~III Pickands.
\newblock Statistical inference using extreme order statistics.
\newblock \emph{Ann. of Stat.}, 3:\penalty0 119--131, 1975.

\bibitem[Pickands(1994)]{pickands1994bayes}
James~III Pickands.
\newblock Bayes quantile estimation and threshold selection for the generalized
  pareto family.
\newblock In \emph{Extreme Value Theory and Applications}, pages 123--138.
  Springer, 1994.

\bibitem[Poirier(2011)]{poirier_bayesian_2011}
Dale~J. Poirier.
\newblock Bayesian interpretations of heteroskedastic consistent covariance
  estimators using the informed {Bayesian} bootstrap.
\newblock \emph{Econometric Reviews}, 30:\penalty0 457--468, 2011.

\bibitem[Politis and Romano(1994)]{politis1994}
DN~Politis and JP~Romano.
\newblock Large sample confidence regions based on subsamples under minimal
  assumptions.
\newblock \emph{The Ann. of Stat.}, 22:\penalty0 2031--2050, 1994.

\bibitem[Romano and Wolf(1999)]{romano1999subsampling}
JP~Romano and M~Wolf.
\newblock Subsampling inference for the mean in the heavy-tailed case.
\newblock \emph{Metrika}, 50:\penalty0 55--69, 1999.

\bibitem[Rubin(1981)]{rubin_bayesian_1981}
Donald Rubin.
\newblock The {Bayesian} {Bootstrap}.
\newblock \emph{The Annals of Statistics}, 9:\penalty0 130--134, 1981.

\bibitem[Scott(2010)]{scott_modern_2010}
Steven~L. Scott.
\newblock A modern {Bayesian} look at the multi-armed bandit.
\newblock \emph{Applied Stochastic Models in Business and Industry},
  26:\penalty0 639--658, 2010.

\bibitem[Smith(1989)]{smith1989extreme}
Richard~L Smith.
\newblock Extreme value analysis of environmental time series: an application
  to trend detection in ground-level ozone.
\newblock \emph{Statistical Science}, 4:\penalty0 367--377, 1989.

\bibitem[Smith(1987)]{smith1987estimating}
RL~Smith.
\newblock Estimating tails of probability distributions.
\newblock \emph{Ann. of Stat.}, 15:\penalty0 1174--1207, 1987.

\bibitem[Taddy et~al.(2015)Taddy, Chen, Yu, and Wyle]{taddy_forests_2015}
Matt Taddy, Chun-Sheng Chen, Jun Yu, and Mitch Wyle.
\newblock Bayesian and empirical {B}ayesian forests.
\newblock In \emph{Proceedings of the 32nd International Conference on Machine
  Learning (ICML 2015)}, 2015.

\bibitem[Taddy et~al.(2016)Taddy, Gardner, Chen, and
  Draper]{taddy_heterogeneous_2015}
Matt Taddy, Matt Gardner, Liyun Chen, and David Draper.
\newblock Nonparametric {B}ayesian analysis of heterogeneous treatment effects
  in digital experimentation.
\newblock \emph{Journal of Business and Economic Statistics}, 2016.
\newblock to appear.

\bibitem[Tierney and Kadane(1986)]{tierney_accurate_1986}
Luke Tierney and Joseph~B. Kadane.
\newblock Accurate approximations for posterior moments and marginal densities.
\newblock \emph{Journal of the American Statistical Association}, 81:\penalty0
  82--86, 1986.

\bibitem[Wyner et~al.(2015)Wyner, Olson, Bleich, and
  Mease]{wyner2015explaining}
Abraham~J Wyner, Matthew Olson, Justin Bleich, and David Mease.
\newblock Explaining the success of adaboost and random forests as
  interpolating classifiers.
\newblock 2015.
\newblock {\it arXiv:1504.07676}.

\end{thebibliography}

\end{document}